\documentclass[a4paper,12pt]{article}
\pdfoutput=1
\usepackage{graphicx,subfigure,amsmath,amssymb}
\usepackage{cite}
\usepackage{color}

\newlength{\dinwidth}
\newlength{\dinmargin}
\begin{document}
\title{\bf  $\bar{B}_{d,s} \to D^{*}_{d,s} V$ and $\bar{B}_{d,s}^* \to D_{d,s} V$ decays in QCD Factorization and Possible Puzzles}
\author{Qin Chang$^{a,b}$, Ling-Xin Chen$^{a}$, Yun-Yun Zhang$^{a}$, Jun-Feng Sun$^{a}$ and Yue-Ling Yang$^{a}$\\
{ $^a$\small Institute of Particle and Nuclear Physics, Henan Normal University, Henan 453007,  China}\\
{ $^b$\small Institute of Particle Physics, Central China Normal University, Wuhan 430079, China}}
\date{}
 \maketitle

\begin{abstract}
Motivated by the rapid development of heavy-flavor experiments, phenomenological studies of nonleptonic $\bar{B}_{d,s} \to D^{*}_{d,s} V$ and $\bar{B}_{d,s}^* \to D_{d,s} V$~($V=\rho\,,K^*$) decays are performed within the framework of  QCD Factorization. Relative to the previous works, the QCD corrections to the transverse amplitudes are evaluated at  next-to-leading order. The theoretical predictions of the observables are updated.  For the measured $\bar{B}_{d,s} \to D^{*}_{d,s} V$ decays, the tensions between theoretical results and experimental measurements, {\it i.e.} ``$R_{ds}^{V}$ puzzle'' and ``$D^{*} V$~(or  $R_{V/\ell\bar{\nu}_\ell}$)  puzzle'', are presented after detailed analyses.  For the $\bar{B}_{d,s}^* \to D_{d,s} V$ decays, they have relatively large branching fractions of the order $\gtrsim{\cal O}(10^{-9})$ and are in the scope of Belle-II and LHCb experiments. Moreover, they also provide a way to crosscheck the possible puzzles  mentioned above through the similar ratios $R_{ds}^{\prime V}$ and $R_{V/\ell\bar{\nu}_\ell}^{\prime}$. More refined  experimental measurements and theoretical efforts are required to confirm or refute such two anomalies.
\end{abstract}


\newpage
\section{Introduction}
Thanks to the efforts of BABAR and Belle collaborations in the past years, most of the $B_{u,d}$ mesons decays with branching fractions $\gtrsim {\cal O}(10^{-7})$ have been measured. With the particle physics entering the LHC era, more rare decays of $B$ mesons, especially of $B_s$ meson, are expected to be well measured. In addition, most recently, the upgrading SuperKEKB/Belle-II experiment has started test operations and succeeded in circulating and storing beams in the electron and positron rings. So, in the near future, the measurements of $B$ meson decays are expected to reach unprecedented precision, which will provide a much more fertile ground for testing the flavor picture of the Standard Model~(SM) and exploring underlying mechanisms. %Correspondingly, the precise theoretical evaluation is essential.  

For the nonleptonic two-body $B$ meson decays, the theoretical evaluation is generally complicated due to the nontrivial QCD dynamics related to the hadronic final states. In order to evaluate the strong interaction corrections to the amplitude, several attractive QCD-inspired approaches, including QCD factorization~(QCDF)~\cite{Beneke1,Beneke2}, the pQCD approach~\cite{KLS1,KLS2} and the soft-collinear effective theory~(SCET)~\cite{scet1,scet2,scet3,scet4}, have been presented. For the case of two light final states, the theoretical evaluations with QCDF approach have been fully developed in the heavy quark limit, for instance Refs.~\cite{Beneke4,Beneke5,Beneke6,Beneke:2005vv,Beneke:2006mk,Beneke:2009ek,du1,du2,Cheng1,Cheng2,Cheng3,Bell:2007tv,Bell:2009nk,Bell:2009fm}. For the case of heavy-light final states, such as $B\to D\pi$ and $D^*\rho$ {\it et al.}, the calculation is generally much more complicated due to the un-negligible $c$ quark mass. In Refs.~\cite{Beneke2,Huber:2016xod}, the factorization formula at two-loop order has been proven, and the explicit results of QCD corrections for $\bar{B}\to DP$, $DV$, $D^*P$ and longitudinally polarized $D^*V$ ($P$ and $V$ are light pseudoscalar and vector mesons, respectively) decays have been presented. 

The $b\to c$ induced nonleptonic $B$ decays, which are tree-dominated and CKM-favored, have relatively large branching fractions and have been widely studied in various theoretical frameworks, for instance Refs.~\cite{Bauer:1986bm, Deandrea:1993ma,Li:2008ts,Azizi:2008ty,Li:2009xf,Chen:2011ut,Fu:2011zzo,Faustov:2012mt,Albertus:2014bfa,Zhou:2015jba}.  In the previous works for the $\bar{B}^0 \to D^{*+} V^-$ decays based on the QCDF, the QCD corrections to the longitudinal amplitude have been fully evaluated at next-to-leading order~(NLO)~\cite{Beneke2} and  next-to-next-to-leading order~(NNLO)~\cite{Huber:2016xod}.
Even though the $\bar{B}^0 \to D^{*+} V^-$ decays are dominated by the longitudinally polarized final states, to keep fairness and consistence, the power-suppressed transverse amplitudes should also be calculated to the same order as the longitudinal one, which is also essential for relatively accurate theoretical results, especially for the polarization fractions.  That is what we would like to do in this paper. Moreover, due to the refined experimental measurements, it is also worth performing detailed phenomenological analyses and testing whether the data and the theoretical results are in agreement. 

In addition to $B$ mesons, their excited states, such as $B^*$ mesons with quantum number of $n^{2s+1}L_J=1^3S_1$ and $J^P=1^-$, also could  decay through the same transitions as $B$ mesons at quark level. Thanks to the rapid development of heavy-flavor experiments~\cite{Abe:2010gxa,Bediaga:2012py,Aaij:2014jba}, even though $B^*$ decays are dominated by the electromagnetic processes $B^*\to B \gamma$, the $B^*$ weak decays with branching fractions $\gtrsim {\cal O}(10^{-9})$ are still hopeful to be observed by Belle-II as analyzed in Refs.~\cite{Chang:2015jla,Chang:2015ead}. Moreover,  owing to the much larger beauty production cross section of $pp$ collisions~\cite{Aaij:2010gn}, the LHC experiments may also provide a lot of experimental information for $B^*$ decays, such as the leptonic $B^*_s\to l^+l^-$ decay with branching fraction $\sim {\cal O}(10^{-11})$ analyzed in Ref.~\cite{Grinstein:2015aua}. Recently, a few interesting theoretical studies of $B^*$ weak decays have been made, for instance Refs.~\cite{Grinstein:2015aua,Wang:2012hu,Zeynali:2014wya,Bashiry:2014qia,Xu:2015eev}. In our previous work~\cite{Chang:2015ead}, the $\bar{B}^{*0} \to D^{*+} V^-$ decays have been studied in the framework of naive factorization~(NF). In this paper, the QCD corrections at NLO will be evaluated with QCDF approach for relatively accurate prediction, and the phenomenological studies will be updated simultaneously. 
 
Our paper is organized as follows. In section 2, the theoretical framework and calculations for $\bar{B}^0_{d,s} \to D^{*+}_{d,s} V^-$ and $\bar{B}_{d,s}^{*0} \to D_{d,s}^+ V^-$ decays are presented with QCDF approach. Section 3 is devoted to the numerical results and discussions. Finally, we give our summary in section 4.

\section{Theoretical Framework and Calculation}
The effective Hamiltonian responsible for the $b\to c \bar{u} q$~($q=d\,,s$) induced  $\bar{B}^0_{d,s} \to D^{*+}_{d,s} V^-$ and $\bar{B}^{*0}_{d,s} \to D^{+}_{d,s} V^-$ decays could be written as
\begin{equation}\label{Heff}
  {\cal H}_{eff}=\frac{G_{F}}{\sqrt{2}}\sum_{q=d,s}V_{cb}V^{*}_{uq}
  \Big\{C_{1}(\mu)Q_{1}(\mu)+C_{2}(\mu)Q_{2}(\mu)\Big\}+h.c.,
\end{equation}
where $G_{F}$ is the Fermi coupling constant, $V_{cb}V^{*}_{uq}$ is the product of CKM matrix elements, and $Q_{1,2}$ are local tree four-quark operators defined as
\begin{eqnarray}
    Q_{1} =
  \bar{c}_{i}{\gamma}_{\mu}(1-{\gamma}_{5})b_{i} 
   \bar{q}_{j} {\gamma}^{\mu}(1-{\gamma}_{5})u_{j} \,,\quad
    Q_{2} =
  \bar{c}_{i}{\gamma}_{\mu}(1-{\gamma}_{5})b_{j} 
  \bar{q}_{j}{\gamma}^{\mu}(1-{\gamma}_{5})u_{i} \,.
    \end{eqnarray}
The corresponding Wilson coefficients $C_{1,2}(\mu)$ summarize the physical contributions above scale of ${\mu}$ and are calculable with the perturbation theory~\cite{Buchalla:1995vs}. 

In order to obtain the decay amplitudes, the remaining work is to  accurately calculate the hadronic matrix elements of every local operators in  effective Hamiltonian. The simplest way to evaluate the hadronic matrix elements is the NF scheme~\cite{Fakirov:1977ta,Cabibbo:1977zv}. However,  in the framework of NF, the amplitudes are renormalization-scale-dependent, and the non-factorizable contributions dominated by the hard gluon exchange are lost. In order to remedy these deficiencies,  the QCDF approach is proposed by BBNS~\cite{Beneke1,Beneke2}. In the framework of QCDF, up to power corrections of order $\Lambda_{\rm QCD}/m_b$, the hadronic matrix elements $\langle M_1M_2| Q_i|\bar{B}^{(*)}\rangle$~($M_1$ is heavy and $M_2$ is light ) obey the factorization formula~\cite{Beneke1,Beneke2}, 
  \begin{eqnarray}
  \langle M_1 M_2|Q_i|\bar{B}^{(*)}\rangle &=&\sum_jF_j^{\bar{B}^{(*)}\to M_1}\int dx{\cal T}_{ij}(x)\Phi_{M_2}(x)\,,
     \label{element}
   \end{eqnarray}
where $F_j^{\bar{B}^{(*)}\to M_1}$ is a $\bar{B}^{(*)}\to M_1$ form factor; $\Phi_{M_2}(x)$ is the light-cone distribution amplitude~(LCDA) for the quark-antiquark Fock state of meson $M_2$; and ${\cal T}_{ij}(x)$ denotes the hard-scattering function, which is calculable order by order from the first principle of perturbative QCD theory. 

Applying the QCDF formula, the amplitude of $\bar{B}\to D^{*+}V^-$ decay could be written as
  \begin{eqnarray}
  {\cal A}_{\lambda}(\bar{B}\to D^{*+}V^-) =  \langle D^{*}V|{\cal H}_{eff}|\bar{B}\rangle=
   \frac{G_{F}}{\sqrt{2}}\, V_{cb} V_{uq}^{\ast}\, \alpha_{1}^{\lambda}\,H_{\lambda}\,,
   \label{amp}
   \end{eqnarray}
where $\lambda=0\,,\pm$ denotes the helicity of $V$ meson; $H_{\lambda}$ is the product of matrix elements of current operators, {\it i.e.},
$H_{\lambda}\equiv \langle V|\bar{q} {\gamma}^{\mu}(1-{\gamma}_{5}) u|0\rangle\langle D^*|\bar{c}{\gamma}_{\mu}(1-{\gamma}_{5})b|\bar{B}\rangle\,$; $\alpha_{1}^{\lambda}$ is the effective flavor coefficient and includes the nonfactorizable contributions. Without the QCD corrections, the NF result, $\alpha_{1}^{\lambda}=C_1+C_2/N_c$, is recovered. The amplitude of $\bar{B}^{*}\to D^{+}V^-$ decay is obtained from the formula  above by replacing $\bar{B}\to \bar{B}^*$ and  $D^*\to D$. The explicit expressions of $H_{\lambda}$ and $\alpha_{1}^{\lambda}$ are given in the following.
   
The decay constant of emitted vector meson is defined through the current matrix element,
\begin{eqnarray}
\langle V(\varepsilon_2,p_2)|\bar{q}\gamma^{\mu} q|0\rangle =-if_Vm_V\varepsilon_{2}^{*\mu}\,,
\end{eqnarray}
where $m_V$ and $\varepsilon_{2}$ denote the mass and the polarization vector, respectively. Meanwhile, with the same conventions as Ref.~\cite{Beneke:2000wa}, the form factors are defined by
\begin{eqnarray}
\langle D^*(\varepsilon_1,p_1)|\bar{c}\gamma_{\mu} b|\bar{B}(p)\rangle &=&
\frac{2iV(q^2)}{m_{B}+m_{D^*}}\epsilon_{\mu\nu\rho\sigma}\varepsilon^{*\nu}_1p^{\rho}p_{1}^{\sigma},\\
\langle D^*(\varepsilon_1,p_1)|\bar{c}\gamma_{\mu}\gamma_5 b|\bar{B}( p)\rangle &=&
2m_{D^*}A_0(q^2)\frac{\varepsilon_1^* \cdot q}{q^2}q_{\mu}
+(m_{D^*}+m_{B})A_1(q^2)\left(\varepsilon^*_{1\mu}-\frac{\varepsilon^*_1\cdot q}{q^2}q_{\mu}\right)\nonumber \\
&&-A_2(q^2)\frac{\varepsilon_1^* \cdot q}{m_{D^*}+m_{B}} \left[(p_1+p)_{\mu}-\frac{m^2_{B}-m^2_{D^*}} {q^2} q_{\mu}\right],
\end{eqnarray}
for $\bar{B}\to D^*$ transition, and  
\begin{eqnarray}
\langle D(p_{1})|\bar{c}\gamma_{\mu} b|\bar{B}^*(\varepsilon, p)\rangle &=&
\frac{2iV(q^2)}{m_{B^*}+m_{D}}\epsilon_{\mu\nu\rho\sigma}\varepsilon^{\nu}p^{\rho}p_{1}^{\sigma},\\
\langle D(p_{1})|\bar{c}\gamma_{\mu}\gamma_5 b|\bar{B}^*(\varepsilon, p)\rangle &=&
2m_{B^*}A_0(q^2)\frac{\varepsilon \cdot q}{q^2}q_{\mu}
+(m_{D}+m_{B^*})A_1(q^2)\left(\varepsilon_{\mu}-\frac{\varepsilon\cdot q}{q^2}q_{\mu}\right)\nonumber \\
&&+A_2(q^2)\frac{\varepsilon \cdot q}{m_{D}+m_{B^*}} \left[(p+p_{1})_{\mu}-\frac{m^2_{B^*}-m^2_{D}} {q^2} q_{\mu}\right],
\end{eqnarray}
for $\bar{B}^*\to D$ transition, where $q=p-p_1=p_2$, $\varepsilon_{(1)}$ is the polarization vector of $\bar{B}^*(D^*)$ meson, and
the sign convention $\epsilon_{0123}=-1$ is taken. Then, after contracting the current matrix elements, we finally obtain
\begin{eqnarray}
\label{eq:H0B}
H_0&=&\frac{i\,f_V}{2\,m_{D^*}}\left[(m_B^2-m_{D^*}^2-m_V^2)(m_B+m_{D^*})A_1^{B\to D^*}(m_V^2)-\frac{4\,m_B^2\,p_c^2}{m_B+m_{D^*}}A_2^{B\to D^*}(m_V^2)\right]\,,\\
\label{eq:HmpB}
H_{\mp}&=&i\,f_V\,m_V\left[(m_B+m_{D^*})A_1(m_V^2)\pm \frac{2\,m_B\,p_c}{m_B+m_{D^*}}V^{B\to D^*}(m_V^2)\right]
\end{eqnarray}
for $\bar{B}\to D^{*+}V^-$ decays, and 
\begin{eqnarray}
\label{eq:H0Bstar}
H_0^{\prime}&=&\frac{i\,f_V}{2\,m_{B^*}}\left[(m_{B^*}^{2}-m_{D}^2+m_V^2)(m_{B^*}+m_{D})A_1^{B^*\to D}(m_V^2)+\frac{4\,m_{B^*}^2\,p_c^{\prime 2}}{m_{B^*}+m_{D}}A_2^{B^*\to D}(m_V^2)\right]\,,\\
\label{eq:HmpBstar}
H_{\mp}^{\prime}&=&-i\,f_V\,m_V\left[(m_{B^*}+m_{D})A_1^{B^*\to D}(m_V^2)\pm \frac{2\,m_{B^*}\,p_c^{\prime}}{m_{B^*}+m_{D}}V^{B^*\to D}(m_V^2)\right]
\end{eqnarray}
for $\bar{B}^{*}\to D^{+}V^-$ decays, in which, 
\begin{eqnarray}
p_c=\frac{\sqrt{[m_B^2-(m_{D^*}+m_V)^2][m_B^2-(m_{D^*}-m_V)^2]}}{2m_B}
\end{eqnarray}
and $p_c^{\prime}$ is obtained from $p_c$ by replacing $m_B\to m_{B^*}$ and $m_{D^*}\to m_D$.

The effective coefficient $\alpha_{1}^{\lambda}$ in the amplitude, Eq.~(\ref{amp}), includes the nonfactorizable contributions from QCD radiative vertex corrections~(the penguin diagrams do not contribute to $\bar{B}^{0}\to D^{*+}V^-$ and $\bar{B}^{*0}\to D^{+}V^-$ decays at the order of $\alpha_s$ ), and could be written as
\begin{equation}\label{eq:alpha1}
\alpha_{1}^{\lambda}= C_{1}^{\rm NLO}+\frac{1}{N_{c}}\,C_{2}^{\rm NLO}+ \frac{{\alpha}_{s}}{4{\pi}}\, \frac{C_{F}}{N_{c}}\,C_{2}^{\rm LO}\, V_1^{\lambda}\,.
  \end{equation}
After calculation, we get the explicit expressions of the vertex corrections $V_1^{\lambda}$ written as
  \begin{eqnarray}\label{eq:v10}
  V_1^0 ={\int}_{0}^{1}du\, {\Phi}_{V}(u)
  \left[ 3\,{\log} \Big( \frac{ m_{b}^{2} }{ {\mu}^{2} } \Big)+ 3\,{\log} \Big( \frac{ m_{c}^{2} }{ {\mu}^{2} } \Big)- 18 +g_0(u)\right]\,,\\
  \label{eq:vmp}
  V_1^{-,+} ={\int}_{0}^{1}du\, \,{\phi}_{b,a}(u)
  \left[ 3\,{\log} \Big( \frac{ m_{b}^{2} }{ {\mu}^{2} } \Big)+ 3\,{\log} \Big( \frac{ m_{c}^{2} }{ {\mu}^{2} } \Big)- 18 +g_{-,+}(u)\right]\,,
  \end{eqnarray}
where ${\Phi}_{V}(u)$ is the leading-twist LCDA and conventionally expanded in Gegenbauer polynomials~\cite{Ball:1998je,BallG},
\begin{eqnarray}
 {\Phi}_{V}(u)=6u\bar{u}\left[ 1+\sum_{n=1}^{\infty}\, \alpha_{n}^{V}(\mu)\,C_{n}^{3/2}(2u-1)\right]\,;
\end{eqnarray}
${\phi}_{a,b}(u)$ are the twist-3 LCDAs given by 
\begin{eqnarray}
\phi_a(u)=\int_{u}^{1}dv\frac{{\Phi}_{V}(v)}{v}\,,    \qquad  
\phi_b(u)=\int_{0}^{u}dv\frac{{\Phi}_{V}(v)}{\bar{v}}\,.
\end{eqnarray}
  It could be found that only the leading-twist LCDA of emitted vector meson contributes to  $V_1^0$ and twist-3 ones contribute to  $V_1^\mp$.
In addition, the loop functions $g_{0,\mp}(u)$ in Eqs.~(\ref{eq:v10}) and (\ref{eq:vmp}) are written as
  \begin{eqnarray}\label{eq:g0}
 g_0(u)&=&
      \frac{c_{a}}{1-c_{a}}\,{\log}(c_{a})-\frac{4\,c_{b}}{1-c_{b}}\,{\log}(c_{b})
      +\frac{c_{d}}{1-c_{d}}\,{\log}(c_{d})-\frac{4\,c_{c}}{1-c_{c}}\,{\log}(c_{c})
      \nonumber \\ &&+
      f(c_{a})-f(c_{b})-f(c_{c})+f(c_{d})
      +2\, {\log}(r_{c}^{2}) \big[ {\log}(c_{a}) -{\log}(c_{b}) \big]-\zeta(r_c)\,,\\
      \label{eq:gT}
 g_{\mp}(u)&=&
      \frac{1+c_{a}}{1-c_{a}}\,{\log}(c_{a})-\frac{4\,c_{b}}{1-c_{b}}\,{\log}(c_{b})
      +\frac{1+c_{d}}{1-c_{d}}\,{\log}(c_{d})-\frac{4\,c_{c}}{1-c_{c}}\,{\log}(c_{c})
         \nonumber \\ &&+
      f(c_{a})-f(c_{b})-f(c_{c})+f(c_{d})
      +2\, {\log}(r_{c}^{2}) \big[ {\log}(c_{a}) -{\log}(c_{b}) \big]-\xi_{\mp}(r_c)\,,
  \end{eqnarray} 
in which, $r_{c} = m_{c}/m_{b}$, $c_{a} = u\,(1-r_{c}^{2})$, $c_{b} = \bar{u}\,(1-r_{c}^{2})$,  $c_{c} = -c_{a}/r_{c}^{2}$, $c_{d} = -c_{b}/r_{c}^{2}$ and
  \begin{eqnarray}
  f(c)=2{\rm Li}_2(\frac{c-1}{c})-\log^2(c)-\frac{2c}{1-c}\log(c)\,.
  \end{eqnarray}
 In the Eqs.~(\ref{eq:g0}) and (\ref{eq:gT}),  the functions $\zeta(r_c)$ and $\xi_{\mp}(r_c)$ contain all of the anti-symmetrical contributions under the transformation $m_c\to -m_c$ (or $r_c\to -r_c$). They are written as 
    \begin{eqnarray}
    \label{zeta}
 \zeta(r_c)&=&-r_{c}\, \Big[ \frac{c_{a}}{(1-c_{a})^{2}}\,{\log}(c_{a})+ \frac{1}{1-c_{a}} \Big]
      -r_{c}^{-1}\,\Big[ \frac{c_{d}}{(1-c_{d})^{2}}\,{\log}(c_{d})+ \frac{1}{1-c_{d}} \Big]\,,  \\
   \label{xi}
\xi_{\mp}(r_c)&=&k_{\mp}\, r_{c}\, \Big[ \frac{2c_{a}-1}{(1-c_{a})^{2}}\,{\log}(c_{a})+ \frac{1}{1-c_{a}} \Big]
      +k_{\mp}\,r_{c}^{-1}\,\Big[ \frac{2c_{d}-1}{(1-c_{d})^{2}}\,{\log}(c_{d})+ \frac{1}{1-c_{d}}\Big]\,,
  \end{eqnarray}
in which, $k_{\mp}\equiv \tilde{H}_{\mp}/H_{\mp}$ with $\tilde{H}_{\mp}=H_{\mp}(A_1\to-A_1)$.  In the limit of $m_c\to 0$,  both $\zeta(r_c)$ and $\xi_{\mp}(r_c)$ vanish, i.e., $\lim\limits_{m_c\rightarrow 0} \zeta(r_c)=\lim\limits_{m_c\rightarrow 0}\xi_{\mp}(r_c)=0$. Moreover,  after taking the limit $m_c\to 0$, one can find that the results of $B\to VV$ decays,  which have been presented in the Eqs. (A.7) and (A.8) of Ref.~\cite{Beneke6}, could be recovered from Eqs. (\ref{eq:v10}), (\ref{eq:vmp}), (\ref{eq:g0}) and (\ref{eq:gT}). 
 
For $g_0(u)$, the only difference between longitudinally polarized $B\to D^*L$ and $B\to DL$ ($L$ is a light meson) decays is the overall sign of $\zeta(r_c)$ (or the sign of $r_{c}$), which has been pointed out in Ref.~\cite{Beneke2} and confirmed in Ref.~\cite{Huber:2016xod}. Such difference could be easily understood from that: (i) after computing the one-loop correction, the ($\bar{q}q^{\prime}$) pair ($q^{(\prime)}$ are light quarks) always retains its $(V-A)$ structure, but the  ($\bar{c}b$) pair has not only $(V-A)$ but also $(V+A)$ structure due to the un-negligible $m_c$ (In this paper,  the contributions of the later are exactly collected into the functions $\zeta(r_c)$ and $\xi_{\mp}(r_c)$, {\it i.e.}, Eqs. (\ref{zeta}) and (\ref{xi})); (ii) only the $V$-current contributes to  $B\to DL$ and only the $A$-current  contributes to  longitudinally polarized $B\to D^*L$. One can also refer to the section 4.4 in Ref.~\cite{Beneke2} for detailed explanation. 
Our result of $V_1^0$ is in consistence with the result in Ref.~\cite{Beneke2}, while the transverse results $V_1^{\mp}$ and $g_{\mp}(u)$ are first presented.  For $g_{\mp}(u)$, because both $V$- and $A$-currents contribute to the transversely polarized $B\to D^*V$ decay, the overall factor $k_{\mp}$ instead of overall sign exists in Eq.~(\ref{xi}). In addition, it should be noted that the strong phase can be obtained by recalling that $r_c^2$ is $r_c^2-i\epsilon$ with $\epsilon>0$ infinitesimal. 

With the amplitudes given above, the branching fraction of $\bar{B}\to D^* V$ decay is defined as 
\begin{eqnarray}\label{br1}
  {\cal B}(\bar{B}\to D^* V)&=&\frac{1}{8\pi}\frac{p_{c}}{m^2_{B}\Gamma_{tot}(\bar{B})}
  \sum_{\lambda}|{\cal A}_{\lambda}(\bar{B}\to D^*V)|^2\,,
\end{eqnarray}
where $\Gamma_{tot}(\bar{B})$ is the total decay width of $\bar{B}$ meson. For $\bar{B}^{*}\to D V$ decays, the definition is obtained from Eq.~(\ref{br1}) by replacing $\bar{B}\to \bar{B}^*$, ${D^*}\to D$, $p_c\to p_c^{\prime}$ and multiplying by an additional factor $1/3$, which is caused by averaging over the spin of initial state $\bar{B}^*$. Besides of the branching fraction, the polarization fractions are also very important observables, which are defined as 
 \begin{equation}
   f_{L,{\parallel},{\perp}} =
   \frac{ {\vert}{\cal A}_{0,{\parallel},{\perp}}{\vert}^{2} }
        { {\vert}{\cal A}_{0}{\vert}^{2}
         +{\vert}{\cal A}_{\parallel}{\vert}^{2}
         +{\vert}{\cal A}_{\perp}{\vert}^{2} }
  \label{eq:pf}\,,
  \end{equation}
where ${\cal A}_{\parallel}$ and ${\cal A}_{\bot}$ are parallel and perpendicular amplitudes, and could be easily gotten through   
${\cal A}_{\parallel,\bot} = ({\cal A}_{-}{\pm}{\cal A}_{+})/\sqrt{2}$.

\section{Numerical Results and Discussions}
%\subsection{Input Parameters}
Before presenting our numerical results, we would like to clarify the input parameters used in the evaluations.  For the CKM matrix elements, we adopt the Wolfenstein  parameterization \cite{Wolfenstein:1983yz} and choose  the parameters $A$ and $\lambda$  as \cite{Charles:2004jd}
  \begin{equation}
  A=0.8227^{+0.0066}_{-0.0136}\,, \quad
  \lambda=0.22543^{+0.00042}_{-0.00031}\,.
%  \bar{\rho}=0.1453^{+0.0133}_{-0.0073}, \quad
 % \bar{\eta}=0.343^{+0.011}_{-0.012}.
  \label{eq:ckminput}
  \end{equation}
For the well-known Fermi coupling constant $G_F$, the masses of mesons and the  total decay widths (or  lifetimes)  of $B$ mesons, we take the central values given by PDG~\cite{PDG14}. However, for $\Gamma_{\rm tot}(B^*)$, there is no available experimental and theoretical information at present. Because the electromagnetic processes $B^*\to B\gamma$ dominate the decays of $B^*$ mesons, we take the approximation $\Gamma_{\rm{tot}}(B^*)\simeq \Gamma(B^*\to B\gamma)$ in our evaluations.  The theoretical predictions for $\Gamma(B^*\to B\gamma)$ have been given in various theoretical models~\cite{Goity:2000dk,Ebert:2002xz,Zhu:1996qy,Aliev:1995wi,Colangelo:1993zq,Choi:2007se,Cheung:2014cka}. In this paper, we take the central values of the latest results~\cite{Choi:2007se,Cheung:2014cka}
\begin{eqnarray} 
 \label{eq:GtotBd}
\Gamma_{\rm{tot}}(B^{*0})&\simeq& \Gamma(B^{*0}\to B^0 \gamma)=(148\pm20)\,{\rm eV},\\
 \label{eq:GtotBs}
\Gamma_{\rm{tot}}(B^{*0}_s)&\simeq& \Gamma(B^{*0}_s\to B^0_s \gamma)=(68\pm17)\,{\rm eV},
\end{eqnarray}
which are in agreement with most of the other theoretical predictions.

As for the light mesons' decay constants and Gegenbauer moments~(at $\mu=2 {\rm GeV}$), we take~\cite{Ball:2006eu,Ball:2007rt}
\begin{eqnarray} 
&&f_{\rho}=216{\pm}3\, {\rm MeV}\,,\quad f_{K^{\ast}}=220{\pm}5\, {\rm MeV}\,,\\
&&a_{1}^{\rho}=0\,,\quad a_{2}^{\rho}=0.10\,,\quad a_{1}^{K^{\ast}}=0.02\,,\quad a_{2}^{K^{\ast}}=0.08\,.
\end{eqnarray}
Then, the residual inputs are the QCD form factors $V(q^2)$ and $A_{1,2}(q^2)$, which are crucial for evaluating the observables of nonleptonic $B^{(*)}$ decays. For the $B\to D^*$ transition, the form factors (or the relevant parameters) could be precisely extracted from the well-measured $B\to D^*\ell\bar{\nu}_\ell$ decay distributions.
After performing a four-dimensional fit to the measurements of exclusive $B\to D^*\ell\bar{\nu}_\ell$ decays,  the HFAG presents the averaged results of the Caprini, Lellouch and Neubert (CLN)~\cite{Caprini:1997mu} form factor parameters for $B\to D^*$ transition~\cite{HFAG}
\begin{eqnarray} \label{cln}
&&h_{A_1}(1)|V_{cb}|=(35.81\pm0.45)\times 10^{-3}\,,\quad \rho^2=1.207\pm0.026\,,\nonumber\\
&& R_1(1)=1.406\pm0.033\,, \quad R_2(1)=0.853\pm0.020\,.
\end{eqnarray}
The QCD form factors $V(q^2)$ and $A_{1,2}(q^2)$ are obtained through the relation~\cite{Caprini:1997mu}
\begin{eqnarray} 
A_1(\omega)=R^*\frac{\omega+1}{2}h_{A_1}(\omega)\,,\quad A_2(\omega)=\frac{R_2(\omega)}{R^*}h_{A_1}(\omega)\,,\quad V=\frac{R_1(\omega)}{R^*}h_{A_1}\,,
\end{eqnarray}
where the ratio $R^*=2\sqrt{m_Bm_{D^*}}/(m_B+m_{D^*})$ and the kinematical variable $\omega=(m_B^2+m_{D^*}^2-q^2)/(2m_Bm_{D^*})$. The $\omega$ dependence of  $h_{A_1}(\omega)$ and  $R_{1,2}(\omega)$ reads~\cite{Caprini:1997mu}
\begin{eqnarray} 
h_{A_1}(\omega)&=&h_{A_1}(1)\left[1-8\rho^2z(\omega)+(53\rho^2-15)z^2(\omega) -(231\rho^2-91)z^3(\omega)\right]\,,\\
R_1(\omega)&=&R_1(1)-0.12(\omega-1)+0.05(\omega-1)^2\,,\\
R_2(\omega)&=&R_2(1)+0.11(\omega-1)-0.06(\omega-1)^2\,,
\end{eqnarray}
with $z(\omega)=(\sqrt{\omega+1}-\sqrt{2})/(\sqrt{\omega+1}+\sqrt{2})$. 
For the form factors of $B_{(s)}^*\to D_{(s)}$ and $B_s\to D^*_s$ transitions, due to the lack of the experimental information, we employ the Bauer-Stech-Wirbel~(BSW) model~\cite{Wirbel:1985ji,Bauer:1988fx} for estimating their values. With the constituent masses $m_u=m_d=0.35\,{\rm GeV}$, $m_s=0.55\,{\rm GeV}$, $m_c=1.7\,{\rm GeV}$, $m_b= 4.9\,{\rm GeV}$ and $w=\sqrt{\langle\vec{p}_{\perp}^2\rangle}=0.4\,{\rm GeV}$, we get 
\begin{eqnarray} 
&&A_1^{B^* \to D}(0)=0.75\,,\quad A_2^{B^* \to D}(0)=0.62\,,\quad V^{B^* \to D}(0)=0.76\,;\\
&&A_1^{B^*_s \to D_s}(0)=0.69\,,\quad A_2^{B^*_s \to D_s}(0)=0.59\,,\quad V^{B^*_s \to D_s}(0)=0.72\,;\\
&&A_1^{B_s \to D^*_s}(0)=0.59\,,\quad A_2^{B_s \to D^*_s}(0)=0.62\,,\quad V^{B_s \to D^*_s}(0)=0.66\,.
\end{eqnarray}
 To be conservative,  $10\%$ uncertainties are assigned to the values above. With the assumption of the nearest pole dominance,  the $q^2$ dependences of form factors read~\cite{Wirbel:1985ji,Bauer:1988fx}
\begin{eqnarray}
A_1(q^2)\simeq\frac{A_1(0)}{1-q^2/m^2_{B_c(1^+)}}\,, \quad
A_2(q^2)\simeq\frac{A_2(0)}{1-q^2/m^2_{B_c(1^+)}}\,, \quad
V(q^2)\simeq\frac{V(0)}{1-q^2/m^2_{B_c(1^-)}},
\end{eqnarray}
where $B_c(J^P)$ is the state of $B_c$ with quantum number of $J^P$~($J$ and $P$ are the quantum numbers of total angular momenta and parity, respectively).

\begin{figure}[t]
\caption{Dependences of the effective tree coefficients $\alpha_1^{\lambda}(D^*\rho)$ at LO and NLO on the renormalization scale $\mu$. }
\begin{center}
\subfigure[]{\includegraphics[scale=0.5]{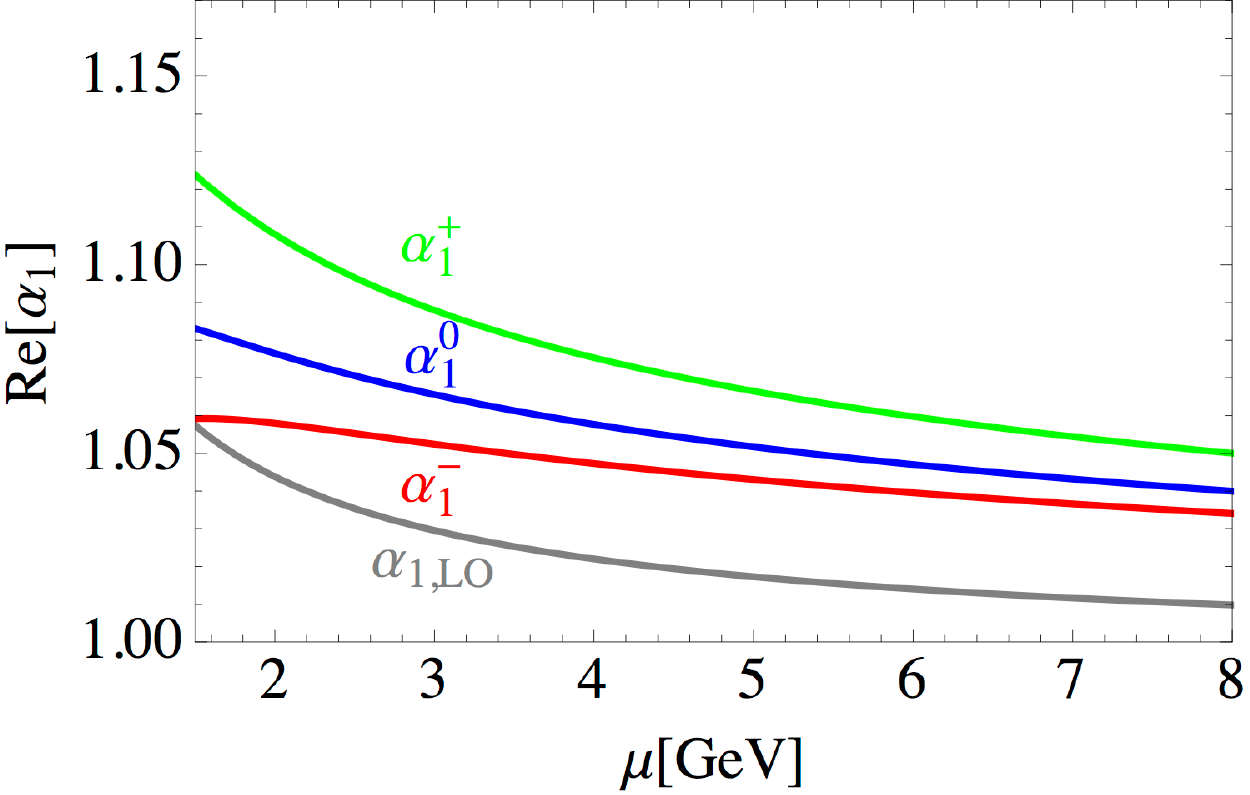}}\qquad\quad
\subfigure[]{\includegraphics[scale=0.515]{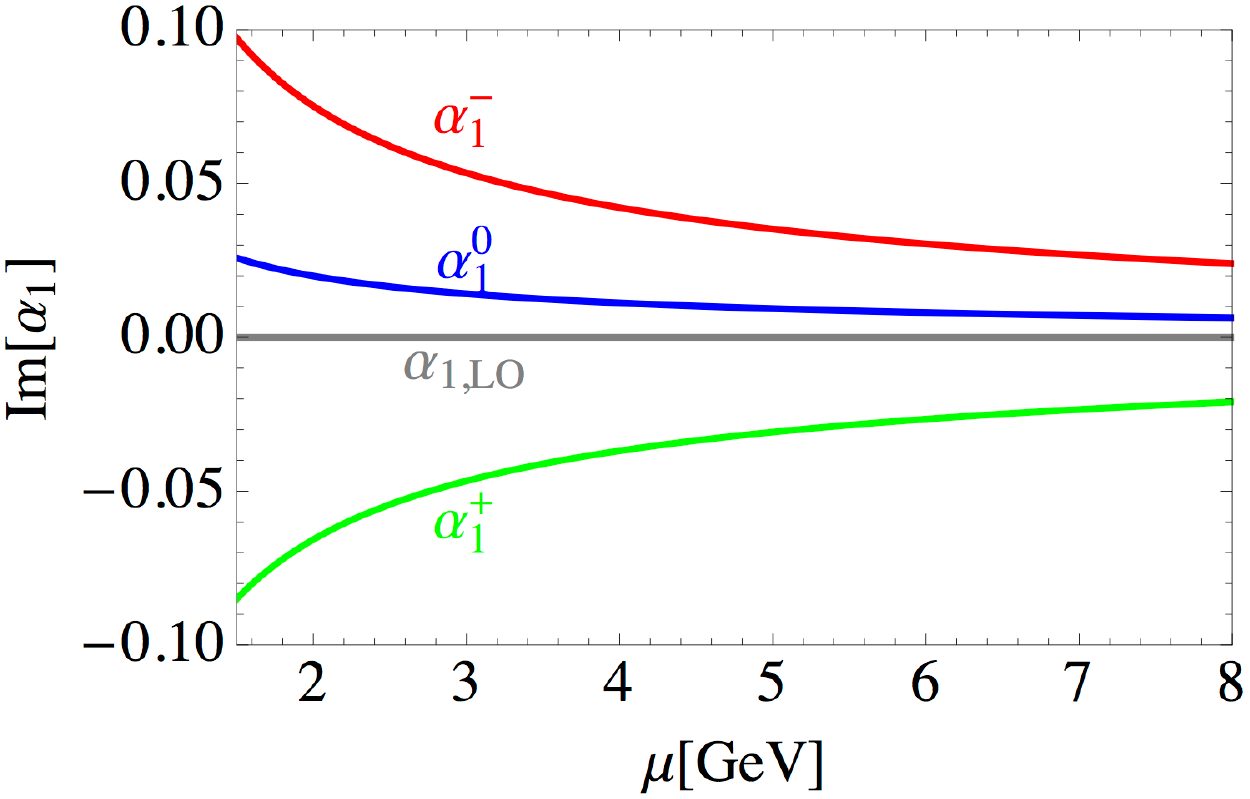}}
\end{center}
\label{fig:sd}
\end{figure}

\begin{table}[t]
\caption{ The  NLO results of effective coefficient $\alpha_1^{\lambda}(D^*\rho)$ at three different  renormalization scales. The LO~(NF) results are also listed for comparison. }
\begin{center}\setlength{\tabcolsep}{5pt}
\begin{tabular}{lcccc}
\hline\hline
                                          &$\mu=m_b/2$   & $\mu=m_b$  & $\mu=2m_b$ \\\hline
$\alpha_1^{0}(D^*\rho)$ & $1.072+0.017i$ & $1.053+0.010i$  &$1.036+0.006i$\\
$\alpha_1^{-}(D^*\rho)$  & $1.056+0.065i$ & $1.044+0.037i$  &$1.031+0.021i$\\
$\alpha_1^{+}(D^*\rho)$ & $1.098-0.056i$& $1.068-0.032i$  &$1.045-0.018i$\\
$\alpha_{1,{\rm LO}}$               &$1.037$& $1.018$  & $1.008$\\
\hline\hline
\end{tabular}
\end{center}
\label{tab:alpha1}
\end{table}

With the theoretical formula and inputs given above, we then present our numerical results and discussions. Within the QCDF framework, the QCD corrections are contained in the effective coefficients $\alpha_i^{\lambda}$, which are generally renormalization-scale-dependent and the dependence is expected to be reduced after the higher order QCD corrections are taken into account.  In Fig.~\ref{fig:sd}, we plot the dependence of tree coefficient $\alpha_1^{\lambda}(D^*\rho)$ on the renormalization scale $\mu$. As Fig.~\ref{fig:sd}~(b) shows, the imaginary part ${\rm Im}[\alpha_1]$, which is zero at LO~(NF result), arises after taking  the NLO corrections into account. Moreover, the sign of ${\rm Im}[\alpha_1^+]$ is different from the ones of  ${\rm Im}[\alpha_1^{0,-}]$. For the real part ${\rm Re}[\alpha_1]$, compared with the LO results, even through the scale dependence has been reduced partly as Fig.~\ref{fig:sd}~(a) shows, the reduction effect is not very significant, which is attributed to that the NLO QCD corrections to $\alpha_1$ associated with small $C_2^{LO}$ is color-suppressed. As found in Refs.~\cite{Beneke:2009ek,Huber:2016xod}, after taking the NNLO correction, which is no longer   color-suppressed, into account, the scale dependence will be significantly improved further. Numerically, the LO and NLO results of coefficient $\alpha_1^{\lambda}(D^*\rho)$ at $\mu=m_b/2\,,m_b\,,2m_b$ are summarized in Table~\ref{tab:alpha1}. Compared with the LO results, it could be found that $|\alpha_1^{0,-,+}(D^*\rho)|$ are enhanced by a factor about $(3.4\,,2.6\,,5.0)\%$ at $\mu=m_b$ by vertex corrections. 

\begin{table}[t]
\caption{The experimental data and theoretical results for the observables of $\bar{B}\to D^* V$ decays. The values listed in the 5-8 columns are the results of the pQCD, the instantaneous Bether-Salpeter method~(BSm), the Heavy quark symmetry~(HQS) and the QCDF with the NNLO corrections to the longitudinal polarization amplitude. }
\begin{center}\footnotesize  \setlength{\tabcolsep}{1pt}
\begin{tabular}{llcccccc}
\hline\hline
Obs. & Decay mode                    &Exp.~\cite{PDG14}         & this work                                  &pQCD~\cite{Li:2008ts}&BSm~\cite{Chen:2011ut,Fu:2011zzo}&HQS~\cite{Deandrea:1993ma} &NNLO~\cite{Huber:2016xod}\\\hline
${\cal B}\,[\times10^{-3}]$
& $\bar{B}^{0} \to D^{*+} K^{*-}$     &$0.33\pm0.06$&$0.58^{+0.04}_{-0.04}$&$0.463^{+0.130+0.101}_{-0.114-0.129}$&$0.64^{+0.07}_{-0.17}$&$0.45$&$0.470^{+0.040}_{-0.039}$\\
&$\bar{B}^{0} \to D^{*+} \rho^-$         &$6.8\pm0.9$  & $10.1^{+0.5}_{-0.5}$     &$7.54^{+2.11+1.58}_{-1.85-1.84}$&$10.3^{+1.7}_{-3.0}$&$8.7$&$9.24^{+0.72}_{-0.71}$\\
&$\bar{B}^{0}_s \to D^{*+}_s K^{*-}$&---                   & $0.54^{+0.20}_{-0.17}$                                     &$0.322^{+0.183+0.098}_{-0.124-0.095}$&$0.56^{+0.06}_{-0.07}$& $0.48$&$0.331^{+0.072}_{-0.067}$\\
&$\bar{B}^{0}_s \to D^{*+}_s \rho^-$  &$9.7^{+2.1}_{-2.2}$~\footnotemark[1]&$9.3^{+3.6}_{-3.1}$  &$5.23^{+2.83+1.77}_{-1.95-1.66}$&$9.0^{+1.5}_{-1.5}$&$8.9$&$6.41^{+1.42}_{-1.31}$\\
\hline $f_L\,[\%]$
& $\bar{B}^{0} \to D^{*+} K^{*-}$     &---                           &$85.5^{+0.4}_{-0.4}$         &$81$&$84.5^{+0.8}_{-0.9}$&---&---\\
&$\bar{B}^{0} \to D^{*+} \rho^-$         &$88.5\pm2.0$         &$88.6^{+0.3}_{-0.3}$  &$85$&$87.8^{+0.7}_{-0.8}$&---&---\\
&$\bar{B}^{0}_s \to D^{*+}_s K^{*-}$&---                       &$84.0^{+2.8}_{-3.9}$  &$83$&$84.1^{+0.4}_{-0.5}$&---&---\\
&$\bar{B}^{0}_s \to D^{*+}_s \rho^-$&$105^{+9}_{-11}$ &$87.3^{+2.4}_{-3.3}$  &$87$&$87.4^{+0.4}_{-0.3}$&---&---\\
\hline
$f_{\|}\,[\%]$
& $\bar{B}^{0} \to D^{*+} K^{*-}$      &---               &$10.8^{+0.3}_{-0.3}$  &--- &$12.8^{+1.2}_{-1.0}$&---&---\\
&$\bar{B}^{0} \to D^{*+} \rho^-$          &---                &$8.4^{+0.2}_{-0.2}$ &--- &$10.1^{+0.9}_{-0.8}$&---&---\\
&$\bar{B}^{0}_s \to D^{*+}_s K^{*-}$ &---                 &$13.2^{+2.7}_{-2.1}$  &--- &$13.3^{+0.5}_{-0.6}$&---&---\\
&$\bar{B}^{0}_s \to D^{*+}_s \rho^-$   &---                &$10.5^{+2.3}_{-1.7}$  &--- &$10.4^{+0.5}_{-0.4}$&---&---\\
\hline\hline
\end{tabular}
\end{center}
\label{tab:B}
\end{table}
  \footnotetext[1]{ The PDG data $9.7^{+2.1}_{-2.2}$ is presented through multiplying the ratio ${\cal B}(\bar{B}^{0}_s \to D^{*+}_s \rho^-)/{\cal B}(\bar{B}^{0}_s \to D^{+}_s \pi^-)=3.2\pm0.6\pm0.3$~(Belle collaboration)~\cite{Louvot:2010rd} by the best value ${\cal B}(\bar{B}^{0}_s \to D^{+}_s \pi^-)=(3.04\pm0.23)\times10^{-3}$.  The Belle's direct measurement gives ${\cal B}(\bar{B}^{0}_s \to D^{*+}_s \rho^-)=(11.9^{+2.2}_{-2.0}\pm1.7\pm1.8) \times10^{-3}$~\cite{Louvot:2010rd}.}

\begin{table}[t]
\caption{The theoretical predictions for the observables of $\bar{B}^*\to D V$ decays.}
\begin{center}\setlength{\tabcolsep}{5pt}
\begin{tabular}{lcccc}
\hline\hline
Decay mode                                   &${\cal B}\,[\times10^{-9}]$ & $f_L\,[\%]$  & $f_{\|}\,[\%]$ \\\hline
$\bar{B}^{*0} \to D^{+} K^{*-}$ & $0.87^{+0.15}_{-0.14}$ &  $84.7^{+0.8}_{-0.9}$  &$12.6^{+0.8}_{-0.9}$\\
$\bar{B}^{*0} \to D^{+} \rho^-$  & $15.1^{+2.5}_{-2.4}$ & $88.0^{+0.7}_{-0.7}$ &$9.9^{+0.7}_{-0.7}$\\
$\bar{B}^{*0}_s \to D^{+}_s K^{*-}$ & $1.66^{+0.28}_{-0.27}$ &$84.9^{+0.8}_{-0.9}$  &$12.4^{+0.8}_{-0.9}$\\
$\bar{B}^{*0}_s \to D^{+}_s \rho^-$   &$28.9^{+4.8}_{-4.5}$& $88.1^{+0.7}_{-0.7}$  & $9.8^{+0.7}_{-0.7}$\\
\hline\hline
\end{tabular}
\end{center}
\label{tab:Bstar}
\end{table}

In Table \ref{tab:B}, we have summarized our numerical results for the observables of $\bar{B}\to D^* V$ decays. Moreover, the experimental data and the results of some previous works are also listed in Table \ref{tab:B} for comparison. In Table \ref{tab:Bstar}, our theoretical predictions for the observables of $\bar{B}^*\to D V$ decays are presented. The theoretical uncertainties of our results in these Tables are obtained through separately evaluating the uncertainty induced by each input parameter and then adding the individual uncertainties in quadrature. The followings are some analyses and discussions.

Using the approximation of $m_V^2\ll m_{B}^2$ and $m_Vm_{D^*}\ll m_{B}^2$, Eqs.~(\ref{eq:H0B}) and (\ref{eq:HmpB}) could be reduced to 
 \begin{eqnarray}
H_0&\approx&\frac{i\,f_V}{2\,m_{D^*}}(m_B^2-m_{D^*}^2)\left[(m_B+m_{D^*})A_1^{B\to D^*}-(m_B-m_{D^*})A_2^{B\to D^*}\right]\,,\\
H_{\mp}&\approx&i\,f_V\,m_V\left[(m_B+m_{D^*})A_1^{B\to D^*}\pm(m_B-m_{D^*})V^{B\to D^*}\right]\,.
\end{eqnarray}
Similarly, for $\bar{B}^*\to DV$ decays, Eqs.~(\ref{eq:H0Bstar}) and (\ref{eq:HmpBstar})  could be simplified as
\begin{eqnarray}
H_0^{\prime}&\approx&\frac{i\,f_V}{2\,m_{B^*}}(m_{B^{*}}^2-m_{D}^2)\left[(m_{B^*}+m_{D})A_1^{B^*\to D}+(m_{B^{*}}-m_{D})A_2^{B^*\to D}\right]\,,\\
H_{\mp}^{\prime}&\approx&-i\,f_V\,m_V\left[(m_{B^*}+m_{D})A_1^{B^*\to D}\pm (m_{B^{*}}-m_{D})V^{B^*\to D}\right]\,.
\end{eqnarray}
From the simplified expression given above, one can obtain the relation $|H_0|:|H_-|:|H_+|\sim 1:2m_V/m_B: 2m_Vm_{D^*}/m_B^2$ for $\bar{B}\to D^{*+}V^-$ decays and $|H_0^{\prime}|:|H_-^{\prime}|:|H_+^{\prime}|\sim 1:2m_V/m_{B^*}: 2m_Vm_{D}/m_{B^*}^2$ for $\bar{B}^{*}\to D^{+}V^-$ decays,  which implies the dominance of the longitudinal polarization in both $\bar{B}\to D^{*+}V^-$ and $\bar{B}^{*}\to D^{+}V^-$  decays. Our numerical results for the polarization fractions listed in Tables \ref{tab:B} and \ref{tab:Bstar}, $f_L\sim [80\%,90\%]$, fulfill such expectation. For the measured $\bar{B}^{0} \to D^{*+} \rho^-$ and $\bar{B}^{0}_s \to D^{*+}_s \rho^-$ decays, our results of  the polarization fractions  are in a good agreement with the data and the theoretical results in previous works. 

For the branching fractions, it could be found from Table  \ref{tab:B} that our results are in consistence with the ones based on the instantaneous Bether-Salpeter method~(BSm)~\cite{Chen:2011ut,Fu:2011zzo}, the Heavy quark symmetry~(HQS)~\cite{Deandrea:1993ma} and the QCDF with the NNLO corrections to the longitudinal polarization amplitude~\cite{Huber:2016xod}, as well as the other theoretical results in Refs.~\cite{Albertus:2014bfa,Faustov:2012mt}, but a bit larger than the result of the pQCD approach~\cite{Li:2008ts}. In addition, the most recent updated pQCD results, ${\cal B}(\bar{B}^{0}_s \to D^{*+}_s \rho^-)=(9.1^{+1.6}_{-1.5})\times 10^{-3}$ and ${\cal B}(\bar{B}^{0}_s \to D^{*+}_s K^{*-})=(0.58\pm0.07)\times 10^{-3}$~\cite{Chen:2011ut} ( the results of $\bar{B}_d$ decays are not updated), agree well with ours. 

Compared with the experimental data, it could be found from Table  \ref{tab:B} that: (i) our QCDF result for ${\cal B}(\bar{B}^{0}_s \to D^{*+}_s \rho^-)$ is in a good agreement with the data; (ii) however, the results~(central values) for ${\cal B}(\bar{B}^{0} \to D^{*+} \rho^-)$ and ${\cal B}(\bar{B}^{0} \to D^{*+} K^{*-})$ are much larger than the data. Moreover, such deviation would be further enlarged if the NNLO QCD corrections, which provide about $4\%$ enhancement to the branching fractions, are included~\cite{Huber:2016xod}. In addition, the results within the other theoretical frameworks  also deviate from the data more or less, which can be seen exactly from Table  \ref{tab:B}. To clarify such possible mismatch, one can define the quantity 
 \begin{equation}\label{eq:rds}
 R_{ds}^{V}\equiv \frac{{\cal B}(\bar{B}_d^{0} \to D^{*+} V^-)}{{\cal B}(\bar{B}^{0}_s \to D^{*+}_s V^-)}\,.
 \end{equation}
Comparing with the $\bar{B}^{0}_s \to D^{*+}_s V^-$~($V=\rho\,,K^*$) decay appeared in denominator, the $\bar{B}_d^{0} \to D^{*+} V^-$ decay appeared in numerator receives additional weak-annihilation corrections, which is however power-suppressed and numerically trivial for the tree-dominated decays~\cite{Beneke2}. So, in the limit of U-spin flavor symmetry, the result $R_{ds}^{V}\simeq1$ is expected. Using the experimental data listed in Table  \ref{tab:B} and the error transfer formula, we get 
 \begin{equation}\label{eq:Rdsexp}
  R_{ds}^{\rho}[{\rm Exp.}]=0.70\pm0.18~(0.57^{+0.17}_{-0.18})\,,
 \end{equation} 
 in which the number in the round brackets is the result gotten by using the direct measurement ${\cal B}(\bar{B}^{0}_s \to D^{*+}_s \rho^-)=(11.9^{+2.2}_{-2.0}\pm1.7\pm1.8) \times10^{-3}$~\cite{Louvot:2010rd} instead of PDG result \footnotemark[1] $(9.7^{+2.1}_{-2.2})\times 10^{-3}$. The results in Eq.~(\ref{eq:Rdsexp}) are much smaller than 1, and imply the significant effect of U-spin flavor-symmetry-breaking.  However, the current theoretical results, 
 \begin{eqnarray}\label{eq:RdsrhoTH}
 R_{ds}^{\rho}[{\rm Theo.}]\simeq1.09\,{\rm(this~work)}\,,1.44\,{\rm(pQCD)}\,,1.14{\rm(BSm)}\,,0.98\,{\rm(HQS)}\,,1.44\,{\rm(NNLO)}
\end{eqnarray}
 are much larger than the experimental data, Eq.~(\ref{eq:Rdsexp}), at about $2.2\,(3.1)\sigma$ \footnotemark[2] \footnotetext[2]{Using the BSW results for the form factors of $B_d\to D^*_d$ transition, $A_1^{B^* \to D}=0.75$, $A_2^{B^* \to D}=0.62$ and $V^{B^* \to D}=0.76$, instead of CLN ones in Eq.~(\ref{cln}), we obtain $R_{ds}^{\rho}[{\rm BSW}]\simeq1.24$, which deviates from the data at the level of $3.0\,(3.9)\sigma$. }, $4.1(5.1)\sigma$, $2.4(3.4)\sigma$, $1.6(2.4)\sigma$ and $4.1(5.1)\sigma$  level, respectively. If the future refined measurements, especially on $\bar{B}^{0}_s \to D^{*+}_s \rho^-$ decay, confirm the large U-spin flavor-symmetry-breaking effects illustrated by Eq.~(\ref{eq:Rdsexp}), it would be a serious challenge to the current theoretical estimation. Besides  $R_{ds}^{\rho}$, the measurements on $R_{ds}^{K^*}$ also could provide a judgment for the possible unexpected large flavor-symmetry-breaking. Numerically, the theoretical results are 
  \begin{eqnarray}\label{eq:RdskTH}
R_{ds}^{K^*}[{\rm Theo.}]\simeq1.07\,{\rm(this~work)}\,,1.44\,{\rm(pQCD)}\,,1.14{\rm(BSm)}\,,0.94\,{\rm(HQS)}\,,1.42{\rm(NNLO)},
\end{eqnarray}
which are very close to the results for $R_{ds}^{\rho}$ given by Eq.~(\ref{eq:RdsrhoTH}), {\it i.e.},  $R_{ds}^{K^*}[{\rm Theo.}]\simeq R_{ds}^{\rho}[{\rm Theo.}]$.
 
 In addition to above-mentioned ``$R_{ds}^{\rho}$ puzzle'', one may find another tension between the theoretical results and the data. Firstly, we would like to emphasize that the values of CLN form factor parameters for $B\to D^*$ transition are purely extracted from the experimental measurements of $B\to D^* \ell \nu_{\ell}$ decays, and therefore, are model-independent. With such inputs, we obtain ${\cal B}(\bar{B}^{0} \to D^{*+} \rho^-)=(10.1^{+0.5}_{-0.5})\times10^{-3}$ and ${\cal B}(\bar{B}^{0} \to D^{*+} K^{*-})=(0.58^{+0.04}_{-0.04})\times10^{-3}$, where the theoretical uncertainties are well controlled due to the precise measurements of $B\to D^* \ell \nu_{\ell}$ decays. However, unfortunately, such theoretical results are about $3.7\sigma$ and $4.2\sigma$, respectively, larger than the data. Due to the fact that the new physics (NP) corrections are generally trivial for the branching fractions of CKM-favored and tree-dominated $\bar{B}^{0} \to D^{*+} V^{-}$ decays, the large gap between the theoretical estimation and experimental data for ${\cal B}(\bar{B}^{0} \to D^{*+} V^-)$, which is called ``$D^{*} V$ puzzle'' in the following discussions for convenience, is hardly to be moderated by the NP contribution. 
 
 In order to check if  the ``$D^{*} V$ puzzle'' is stable, an improved way is to perform an measurement on the ratio defined by
  \begin{eqnarray}\label{eq:rvl}
 R_{V/\ell\bar{\nu}_\ell}\equiv \frac{\Gamma(\bar{B}^{0} \to D^{*+} V^{-} )}{\left. d\Gamma(\bar{B}\to D^*\ell \bar{\nu}_{\ell})/dq^2\right|_{q^2= m_V^2}}\,,
\end{eqnarray} 
% \begin{eqnarray}
% R_{V/lv}^{\rho}\equiv \frac{\frac{d\Gamma(B\to D^*\ell \bar{\nu}_{\ell})}{dq^2}|_{q^2=m_V^2}}{\Gamma(\bar{B}^{0} \to D^{*+} V^{-} )}\,,
%\end{eqnarray} 
in which $\ell=e\,,\mu$.
Firstly, we would like to estimate its theoretical result.  Neglecting the lepton mass, the differential decay rate can be written as
% \begin{eqnarray}\label{eq:SdG}
%\frac{d\Gamma(\bar{B}\to D^*\ell \bar{\nu}_{\ell})}{dq^2}=\frac{G_F^2|V_{cb}|^2|\vec{p}|q^2}{96\pi^3m_{B}^2}\,(1-\frac{m_\ell^2}{q^2})^2\,\left[(H_{++}^2+H_{--}^2+H_{00}^2)(1+\frac{m_\ell^2}{2\,q^2})+\frac{3m_\ell^2}{2q^2}H_{0t}^2\right]\,.
% \end{eqnarray}
 \begin{eqnarray}\label{eq:SdG}
\frac{d\Gamma(\bar{B}\to D^*\ell \bar{\nu}_{\ell})}{dq^2}=\frac{G_F^2|V_{cb}|^2|\vec{p}|q^2}{96\pi^3m_{B}^2}\,\left(H_{++}^2+H_{--}^2+H_{00}^2\right)\,,
\end{eqnarray}
where $H_{00,\pm\pm}$ are the helicity amplitudes and are $q^2$-dependent. One may refer to Refs.~\cite{Korner:1987kd, Korner:1989qb,Fajfer:2012vx,Celis:2012dk} {\it et al.} for the details. At $q^2= m_V^2$, we get
%For the case of $\ell=e\,,\mu$ and at $q^2\simeq m_V^2$, the terms related to $m_\ell^2/q^2$ can be safely neglected, and moreover, 
 \begin{eqnarray}
 \left.|\vec{p}|\right|_{q^2= m_V^2}=p_c\,\quad{\rm and} \quad  \left. |H_{00,--,++}|^2\right|_{q^2= m_V^2}=\frac{|H_{0,-,+}|^2}{(f_Vm_V)^2}\,,
  \end{eqnarray}
 where $H_{0,-,+}$ have been given by Eqs.~(\ref{eq:H0B}) and (\ref{eq:HmpB}), and the explicit expression for $H_{00,--,++}$ could be found in Ref.~\cite{Celis:2012dk}. Then, further considering that $|\alpha_1^{0}|^2 \approx |\alpha_1^{-}|^2 \approx |\alpha_1^{+}|^2$ numerically, we finally obtain 
  \begin{eqnarray}\label{eq:rvlsim}
 R_{V/\ell\bar{\nu}_\ell}\simeq6\pi^2|V_{uq}^*|^2|\alpha_1|^2f_V^2\left.\frac{m_V^2}{q^2}\right|_{q^2= m_V^2}= 6\pi^2f_V^2|V_{uq}|^2|\alpha_1|^2\,,
    \end{eqnarray}
 which is independent of the form factors, and principally could be precisely determined. Numerically, for $\bar{B}^{0} \to D^{*+} \rho^-$ and $\bar{B}^{0} \to D^{*+} K^{*-}$ decays, one can easily get the theoretical prediction
 \begin{eqnarray}\label{eq:Rvlth1}
 R_{\rho/\ell\bar{\nu}_\ell}{\rm [Theo.]}&\simeq&  2.91\,{\rm GeV^2}\cdot\left[\frac{ f_\rho}{0.216\,{\rm GeV}}\right]^2\left[\frac{|V_{ud}|}{0.9743}\right]^2\left[\frac{|\alpha_1|}{|1.053 + 0.010i|}\right]^2\,,\\
 \label{eq:Rvlth2}
 R_{K^*/\ell\bar{\nu}_\ell}{\rm [Theo.]}&\simeq&  0.16\,{\rm GeV^2}\cdot\left[\frac{ f_{K^*}}{0.220\,{\rm GeV}}\right]^2\left[\frac{|V_{us}|}{0.2254}\right]^2\left[\frac{|\alpha_1|}{|1.053 + 0.010i|}\right]^2\,.
  \end{eqnarray}
Using the most recent result $|\alpha_1|=1.07$ at NNLO in QCDF~\cite{Huber:2016xod}, one can obtain the similar results $ R_{\rho/\ell\bar{\nu}_\ell}^{\rm NNLO}=3.01$ and  $R_{K^*/\ell\bar{\nu}_\ell}^{\rm NNLO}=0.17$. Then, using the distribution of $d\Gamma(\bar{B}\to D^*\ell \bar{\nu}_{\ell})/d\omega$ measured by Belle collaboration~\cite{Dungel:2010uk} and the data of ${\cal B}(\bar{B}^{0} \to D^{*+} \rho^-)$ and ${\cal B}(\bar{B}^{0} \to D^{*+} K^{*-})$ listed in Table~\ref{tab:B}, we obtain the experimental results
 \begin{eqnarray}\label{eq:Rvlexp1}
  R_{\rho/\ell\bar{\nu}_\ell}{\rm[Exp.]}&=&2.13\pm0.34\,,\\
  \label{eq:Rvlexp2}
  R_{K^*/\ell\bar{\nu}_\ell}{\rm[Exp.]}&=&0.10\pm0.02\,.
    \end{eqnarray}
In the estimation, to get the differential decay rate at $q^2= m_V^2$, we pick out the measurement at the bin $\omega\in[1.45,1.50]$, which covers the point $q^2= m_{\rho,K^*}^2$,  and take the approximation $d\Gamma(\bar{B}\to D^*\ell \bar{\nu}_{\ell})/d\omega\left.\right|_{q^2= m_V^2}\simeq\frac{\Delta \Gamma(\bar{B}\to D^*\ell \bar{\nu}_{\ell}) \left.\right|_{[1.45,1.50]}}{1.50-1.45}$. Comparing Eqs. (\ref{eq:Rvlth1}) and (\ref{eq:Rvlth2}) with Eqs.~(\ref{eq:Rvlexp1}) and (\ref{eq:Rvlexp2}), one may find that the SM expectations for  $R_{\rho/\ell\bar{\nu}_\ell}$ and $R_{K^*/\ell\bar{\nu}_\ell}$ deviate from the experimental results by about $2.3\sigma$ and $3.0\sigma$, respectively. In Ref.~\cite{Huber:2016xod}, the authors have also pointed out that the the deviation is at the level of $2-3\sigma$ to the NNLO accuracy. A specific experimental measurement or analysis on $R_{V/\ell\bar{\nu}_\ell}$ at a very narrow bin covering  $q^2= m_{\rho,K^*}^2$ is required for a much more reliable result.

For $\bar{B}^{*} \to D^{+} V^{-}$ decays, it could be found from Table \ref{tab:Bstar} that all decays have the branching fractions $\gtrsim {\cal O}(10^{-9})$, and therefore, are hopeful to be measured by Belle-II experiment, which have been pointed out in Ref.~\cite{Chang:2015ead}. Moreover, they are also in the scope of LHCb experiment, which can be seen from the following analysis. To make an estimate, we take  $\bar{B}^{0}\to D^+\pi^-$ decay as a reference.  Using the data corresponding to integrated luminosities of $1.0fb^{-1}$ of $pp$ collisions at a center-of-mass energy of $\sqrt{s}=7{\rm TeV}$, about $1.06\times 10^{5}$ $\bar{B}^{0}\to D^+\pi^-$ decay events have been measured by LHCb collaboration~\cite{Aaij:2013qqa}. 
After high-luminosity upgrade, a data sample of $50\,{\rm fb}^{-1}$ will be collected  by LHCb collaboration at a much higher $\sqrt{s}=14\,{\rm TeV}$, which results in a further enhancement of $b\bar{b}$ production by a factor about 2~\cite{Bediaga:2012py,LHCb:upgrade}. Moreover, the most of $B$ mesons detected at LHC are mainly produced through $B^*\to B\gamma$ decays because $B^*$ mesons are often produced by about 3 times more than the $B$ mesons~\cite{Buskulic:1995mt}. Finally, further considering $\frac{{\cal B}(\bar{B}^{*0} \to D^{+} \rho^-)}{{\cal B}(\bar{B}^{0}\to D^+\pi^-)}= \frac{15.1\times10^{-9}}{26.8\times10^{-4}}\simeq 0.56\times10^{-5}$, we estimate that of the order of ${\cal O}(100)$ $\bar{B}^{*0} \to D^{+} \rho^-$ decay events could be collected by LHCb. Furthermore, taking $\bar{B}^{0}_s\to D^+_s\pi^-$ as reference instead of $\bar{B}^{0}\to D^+\pi^-$ and revisiting the estimate above, one may find about ${\cal O}(10)$ $\bar{B}^{*0}_s \to D^{+}_s \rho^-$ decay events are expected to be observed in the high-luminosity LHC era.

The $\bar{B}^{*0} \to D^{+} V^-$ decays occur with the same transition as  $\bar{B}^{0} \to D^{*+} V^-$ decays at quark level, and have similar amplitudes as $\bar{B}^{0} \to D^{*+} V^-$ decays. So,  once measured in the future, they are expected to provide crosschecking of the possible  $R_{ds}^{V}$ and $D^{*} V$~(or  $R_{V/\ell\bar{\nu}_\ell}$)  puzzles mentioned above. For this point, one can define the ratios    
 \begin{equation}
 R_{ds}^{\prime V}\equiv \frac{{\cal B}(\bar{B}_d^{*0} \to D^{+} V^-)}{{\cal B}(\bar{B}^{*0}_s \to D^{+}_s V^-)}\,, 
 \quad R^{\prime}_{V/\ell\bar{\nu}_\ell}\equiv \frac{\Gamma(\bar{B}^{*0} \to D^{+} V^{-} )}{d\Gamma(\bar{B}^*\to D\ell \bar{\nu}_{\ell})/dq^2|_{q^2= m_V^2}}\,,
 \end{equation}
which are similar to the ones for $B$ decays, Eqs.~(\ref{eq:rds}) and (\ref{eq:rvl}). For $R^{\prime}_{V/\ell\bar{\nu}_\ell}$, it is expected to be equal to $R_{V/\ell\bar{\nu}_\ell}$, Eq.~(\ref{eq:rvlsim}). For $R_{ds}^{\prime V}$, numerically, we get
 \begin{equation}
 R_{ds}^{\prime \rho}[{\rm theo.}]=\frac{\tau_{B_s^*}}{\tau_{B^*}}\cdot1.12\simeq0.52\,,\quad  R_{ds}^{\prime K^*}[{\rm theo.}]=\frac{\tau_{B_s^*}}{\tau_{B^*}}\cdot1.12\simeq0.52\,.
  \end{equation}
Comparing  $R^{\prime V}_{ds}$ with $R_{ds}^{V}$, one can find that the main difference is induced by the lifetimes of $B_{d,s}$ and $B_{d,s}^*$\,, {\it i.e.}, $\tau_{B_s}/\tau_{B_d}\approx 1$ but $\tau_{B_{s}^*}/\tau_{B_{d}^*}\approx 0.46$~(see Eqs.(\ref{eq:GtotBd}) and (\ref{eq:GtotBs})). All of the findings  above for $B_{d,s}^*$ decays are awaiting the future experimental test. 

\section{Summary}
Motivated by the rapid development of heavy-flavor experiments, we perform phenomenological studies of nonleptonic $\bar{B}_{d,s} \to D^{*}_{d,s} V$ and $\bar{B}_{d,s}^* \to D_{d,s} V$~($V=\rho\,,K^*$) decays in detail. The amplitudes are  calculated carefully in the framework of QCD Factorization, in which, relative to the previous works, the NLO QCD corrections to the transverse amplitudes are evaluated. Our theoretical results for the branching fractions and polarization fractions are summarized in Tables~\ref{tab:B} and \ref{tab:Bstar}. After detailed analyses, two possible puzzles relevant to $\bar{B}_{d,s} \to D^{*}_{d,s} V$ decays are presented, which are: (i) the current theoretical results for the ratio $ R_{ds}^{V}$, which reflects the effects of flavor-symmetry-breaking and is defined by Eq.~(\ref{eq:rds}), tend to $\gtrsim 1$ and significantly conflict with the experimental results $0.70\pm0.18~(0.57^{+0.17}_{-0.18})$; (ii) With the form factors extracted  precisely from semileptonic $\bar{B}\to D^*\ell\bar{\nu}_\ell$ decays, the QCDF results for ${\cal B}(\bar{B}^{0} \to D^{*+} \rho^-)$ and ${\cal B}(\bar{B}^{0} \to D^{*+} K^{*-})$ deviate from the data by about $3.7\sigma$ and $4.2\sigma$, respectively. Such tension is also reflected by the form-factor-independent ratio $R_{V/\ell\bar{\nu}_\ell}$ defined by Eq.~(\ref{eq:rvl}), which could be well determined in the SM. The more experimental and theoretical efforts are required to confirm or refute such two anomalies.  For the $\bar{B}_{d,s}^* \to D_{d,s} V$ decays, they have relatively large branching fractions of the order $\gtrsim{\cal O}(10^{-9})$ and are hopeful to be measured by the Belle-II and LHCb experiments. Moreover, they also provide a way for crosschecking of above-mentioned  ``$R_{ds}^{V}$ and $D^{*} V$~(or  $R_{V/\ell\bar{\nu}_\ell}$)   puzzles'' through the similar ratios $R_{ds}^{\prime V}$ and $R_{V/\ell\bar{\nu}_\ell}^{\prime}$. All of the findings in this paper are awaiting the precise test by the refined measurements at LHC and SuperKEKB/Belle-II in the near future.

\section*{Acknowledgments}
  We thank   De-Shan Yang at UCAS, Ya-Dong Yang and Xin-Qiang Li at CCNU for helpful suggestions and discussions.
  This work is supported by the National Natural Science Foundation of China (Grant Nos. 11547014, 11475055 and 11275057).
  Q. Chang is also supported by the Foundation  for the Author of National Excellent Doctoral  Dissertation of P. R. China (Grant No. 201317),  the Program for Science and Technology Innovation Talents in Universities of Henan Province  (Grant No. 14HASTIT036).

%%%%%%%%%%%%%%%%

\end{document}